\pdfoutput=1 
\documentclass[10pt,conference]{IEEEtran}
\usepackage[bottom]{footmisc}
\usepackage{tabularx,array,siunitx}
\usepackage{amsmath,amssymb}
\usepackage{float}
\usepackage{multirow}
\usepackage{graphicx}
\usepackage{color}
\usepackage[table,x11names]{xcolor}
\usepackage{tabularray}
\usepackage{booktabs}
\usepackage{soul}
\usepackage[
    ruled,
    lined,
    shortend,
    linesnumbered,
]{algorithm2e}
\usepackage[capitalise]{cleveref}
\usepackage{cite} 
\usepackage{threeparttable}
\usepackage{makecell}
\usepackage{listings}
\usepackage{glossaries}
\usepackage{siunitx}
\usepackage{enumitem}
\usepackage{caption}
\usepackage{subcaption} 
\usepackage{balance}
\usepackage{tikz}       
\usepackage{pgf-pie}    
\usepackage{pgfplots}
\usepackage{pgfplotstable}
\pgfplotsset{compat=1.18}
\usetikzlibrary{pgfplots.groupplots}
\usetikzlibrary{shapes.geometric, arrows}
\usetikzlibrary{positioning}

\pgfplotsset{
    /pgfplots/ybar legend/.style={
        /pgfplots/legend image code/.code={%
        \draw[##1,/tikz/.cd, bar width=3pt, yshift=-0.4em, bar shift=0pt]
                plot coordinates {(0cm,0.8em)};},
    },
}

\newacronym{ml}{ML}{Machine Learning}
\newacronym{dt}{DT}{Decision Tree}
\newacronym{svm}{SVM}{Support Vector Machine}
\newacronym{nn}{NN}{Neural Network}
\newacronym{bnn}{BNN}{Binary Neural Network}
\newacronym{tnn}{TNN}{Ternary Neural Network}
\newacronym{mlp}{MLP}{Multi-Layer Perceptron}
\newacronym{cnn}{CNN}{Convolutional Neural Network}
\newacronym{rf}{RF}{Random Forest}
\newacronym{snn}{SNN}{Spiking Neural Network}
\newacronym{knn}{k-NN}{k-nearest neighbors}
\newacronym{fe}{FE}{Flexible Electronics}
\newacronym{adc}{ADC}{Analog-to-Digital Converter}
\newacronym{dac}{DAC}{Digital-to-Analog Converter}
\newacronym{eeg}{EEG}{electroencephalography}
\newacronym{ibi}{IBI}{Interbeat Interval}
\newacronym{rr}{RR}{Respiration Rate}
\newacronym{eda}{EDA}{electrodermal activity}
\newacronym{emg}{EMG}{electromyography}
\newacronym{ecg}{ECG}{electrocardiogram}
\newacronym{resp}{RESP}{respiration}
\newacronym{acc}{ACC}{accelerometer}
\newacronym{bvp}{BVP}{blood volume pulse}
\newacronym{temp}{TEMP}{temperature}
\newacronym{hr}{HR}{heart rate}
\newacronym{ppg}{PPG}{photoplethysmogram}
\newacronym{gsr}{GSR}{galvanic skin response}
\newacronym{flexic}{FlexIC}{Flexible Integrated Circuit}
\newacronym{tft}{TFT}{Thin-Film Transistor}
\newacronym{rom}{ROM}{Read-Only Memory}
\newacronym{nsga2}{NSGA-II}{Non-dominated Sorting Genetic Algorithm}
\newacronym{lut}{LUT}{Look-Up Table}
\newacronym{ppa}{PPA}{power, performance and area}
\newacronym{pdk}{PDK}{Process-Design Kit}
\newacronym{wesad}{WESAD}{Wearable Stress and Affect Detection}
\newacronym{fft}{FFT}{Fast Fourier Transform}
\newacronym{iot}{IoT}{Internet-of-Things}
\newacronym{dse}{DSE}{Design Space Exploration}
\newacronym{nre}{NRE}{non-recurring engineering}
\newacronym{ovr}{OvR}{One-vs-Rest}
\newacronym{relu}{ReLU}{Rectified Linear Unit}
\newacronym{igzo}{IGZO}{Indium Gallium Zinc Oxide}
\newacronym{nas}{NAS}{Neural Architecture Search}
\newacronym{ltp}{LTP}{Lottery-Ticket Pruning}

\definecolor{colordt}{RGB}{255,0,0}        
\definecolor{colormlp}{RGB}{255,255,0}     
\definecolor{colorsvm}{RGB}{0,0,255}       
\definecolor{colorbnn}{rgb}{0.29,0.33,0.13} 
\definecolor{colortnn}{rgb}{0.43,0.21,0.1} 

\definecolor{areaADC}{RGB}{251,128,114}    
\definecolor{areaFEx}{RGB}{158,202,225}     
\definecolor{areaClf}{RGB}{77,175,74}     
\definecolor{areaSystem}{RGB}{152,78,163} 

\definecolor{powerADC}{RGB}{0,153,0}     
\definecolor{powerFEx}{RGB}{102,190,102} 
\definecolor{powerClf}{RGB}{190,255,190} 

\definecolor{myarrowcolor}{RGB}{0,0,0}

\lstdefinestyle{mystyle}{
    backgroundcolor=\color{backcolour},   
    commentstyle=\color{codegreen},        
    keywordstyle=\color{blue},             
    numberstyle=\tiny\color{codegray},     
    stringstyle=\color{codepurple},        
    basicstyle=\ttfamily\small,            
    breakatwhitespace=false,               
    breaklines=true,                       
    captionpos=b,                          
    keepspaces=true,                       
    numbers=left,                          
    numbersep=5pt,                         
    showspaces=false,                      
    showstringspaces=false,                
    showtabs=false,                        
    tabsize=2,                             
    language=Python                        
}
\newcommand{\blue}[1]{{\color{black}#1}}

\def\BibTeX{{\rm B\kern-.05em{\sc i\kern-.025em b}\kern-.08em
    T\kern-.1667em\lower.7ex\hbox{E}\kern-.125emX}}

\makeatletter
\newcommand*\titleheader[1]{\gdef\@titleheader{#1}}
\AtBeginDocument{%
  \let\st@red@title\@title
  \def\@title{%
    \bgroup\normalfont\normalsize\centering\@titleheader\par\egroup
    \vskip1ex\st@red@title}
}
\makeatother

\title{Invited Paper: Feature-to-Classifier Co-Design \\for Mixed-Signal Smart Flexible Wearables \\for Healthcare at the Extreme Edge\vspace{-1ex}}

\author{
    \IEEEauthorblockN{
        Maha Shatta\IEEEauthorrefmark{1}, 
        Konstantinos Balaskas\IEEEauthorrefmark{4}, 
        Paula Carolina Lozano Duarte\IEEEauthorrefmark{1},
        Georgios Panagopoulos\IEEEauthorrefmark{9},\\
        Mehdi B. Tahoori\IEEEauthorrefmark{1},
        Georgios Zervakis\IEEEauthorrefmark{4}
    }
    
    \IEEEauthorblockA{
    \IEEEauthorrefmark{1}Karlsruhe Institute of Technology, DE,
    \IEEEauthorrefmark{4}University of Patras, GR,
    \IEEEauthorrefmark{9}National Technical University of Athens, GR
    }

    \IEEEauthorblockA{
    \IEEEauthorrefmark{1}\{maha.shatta, paula.duarte, mehdi.tahoori\}@kit.edu,
    \IEEEauthorrefmark{4}\{kompalas, zervakis\}@ceid.upatras.gr, 
    \IEEEauthorrefmark{9}gepanago@mail.ntua.gr   \\
    }
\vspace{-5ex}
}
\titleheader{\vspace{-25pt}Accepted at 2025 IEEE/ACM International Conf. on Computer-Aided Design (ICCAD) Oct. 26-30 2025, Munich, DE}

\begin{document}
\bstctlcite{IEEEexample:BSTcontrol} 
\maketitle
\begin{abstract}
Flexible Electronics (FE) offer a promising alternative to rigid silicon-based hardware for wearable healthcare devices, enabling lightweight, conformable, and low-cost systems.
However, their limited integration density and large feature sizes impose strict area and power constraints, making \blue{ML-based healthcare systems--integrating analog frontend, feature extraction and classifier--}particularly challenging.
Existing FE solutions often \blue{neglect potential system-wide solutions and} focus on the classifier, overlooking the substantial hardware cost of feature extraction and Analog-to-Digital Converters (ADCs)--both major contributors to area and power consumption.
In this work, we present a holistic mixed-signal feature-to-classifier co-design framework for flexible \blue{smart wearable} systems.
To the best of our knowledge, we design the first analog feature extractors in FE, significantly reducing feature extraction cost.
We further propose an hardware-aware NAS-inspired feature selection strategy within ML training, enabling efficient, application-specific designs.
Our evaluation on healthcare benchmarks shows our approach delivers highly accurate, ultra-area-efficient flexible systems--ideal for disposable, low-power wearable monitoring.
\end{abstract}


\glsresetall
\section{Introduction}
\label{sec:introduction}

In recent years, the demand for advanced healthcare applications has grown significantly, driven by the increasing need for continuous and personalized monitoring of an individual's physiological state~\cite{khan2024anomaly}.
As healthcare systems shift from reactive to proactive models, the ability to continuously monitor patients outside clinical settings has become increasingly important, with a growing interest towards wearable devices~\cite{mahato2024hybrid,jeong2025exploiting,Schmidt2018IntroducingWA:wesad}.
These devices capture physiological signals (e.g., \gls{eda}) during everyday activities, enabling real-time, healthcare tracking at the extreme edge.

Significant efforts have focused on the algorithmic development of wearable devices, particularly with the use of \gls{ml} algorithms~\cite{Arsalan2019ClassificationOP, Kumar2021HierarchicalDN,Aqajari2020GSRAF, Shiyi:IoT2022:StressMonitoring}.
Targeting tasks such as stress, heart or respiration monitoring~\cite{MIT:stressDataset:affectiveROAD,logacjov2021harth,bhattacharya2023coswara}, works span a wide range of \gls{ml} models
which analyze diverse biosignals for real-time health monitoring~\cite{Arsalan2019ClassificationOP, Kumar2021HierarchicalDN, Aqajari2020GSRAF, tazarv:2021:personalized, boateng:mit2016:stressaware, golgouneh:springer2020:rigitProcessorStrees, attaran:2018IEEE:rigitStress}.
However, little attention has been paid to the underlying hardware, or to potential co-design opportunities, \blue{since the sensing/analog frontend is typically designed in isolation from the \gls{ml} classifier.}
Most commercial wearables still rely on rigid, silicon-based microcontrollers, which hinder comfort and skin conformability; their general-purpose design wastes energy, reducing battery life, while high manufacturing costs limit affordability and adoption~\cite{mishra:2020:commodityHardware}.


\gls{fe} have emerged as a compelling alternative to rigid silicon-based platforms for wearable healthcare.
Built on lightweight, conformable substrates, \glspl{flexic} conform to body contours, improving comfort and long-term wearability~\cite{Gao:Nature2022:FlexibleSensor,Heng:AM2022:FlexHumanMachInterfaces}.
\gls{fe} also enable ultra–low-cost, disposable systems for clinical and consumer use.
Their fabrication sidesteps many silicon constraints (cleanrooms, packaging), enabling fast, low-cost--even portable--production~\cite{Bleier:ISCA:2020:printedmicro,flexicores,ozer:nature2024:bendableRiscV}, and offers environmental gains (lower water/energy use and carbon footprint) over silicon processes~\cite{ozer:nature2024:bendableRiscV}.
Together, these features position \gls{fe} as a sustainable, scalable, and user-friendly foundation for next-generation accessible wearable health monitoring systems.

The integration of flexible components into wearables mainly targets mechanically bendable bio-sensors, capable of conformably acquiring diverse signals in real time (e.g., skin temperature)~\cite{Wang2017FlexibleSE,Gao:Nature2022:FlexibleSensor,Yang:RSC2019:flexmonitoring,yoon:nature2016:flexibleSensorPatch}.
On the algorithmic side, \glspl{flexic} have been used for implementing \gls{ml} classification~\cite{Ozer:Nature:2020,afentaki2025islped}, even targeting applications such as malodour detection~\cite{ozer2023malodour}.
However, \gls{fe} are constrained by large feature sizes and limited integration density--stemming from their low-cost fabrication processes~\cite{tahoori2025computing,Henkel:ICCAD2022:expedition}--which lead to increased area and power requirements, making it challenging to implement complex flexible \gls{ml}-based systems.
\blue{Furthermore, integrating FE circuits with commercial off-the-shelf silicon components can be cumbersome--e.g., for the analog front end--necessitating co-design and co-fabrication of the entire system, from sensor and analog interface to classifier, on the same substrate.}

\begin{figure}[t!]
    \centering
    \begin{subfigure}{0.8\columnwidth}
        \caption{\hfill \break\vspace{-3ex}}
        \centering
        \includegraphics[width=\columnwidth]{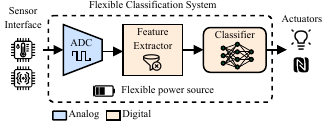}\vspace{-4ex}
        \label{fig:system_digital}
    \end{subfigure}
    \begin{subfigure}{0.8\columnwidth}
        \caption{\hfill \break\vspace{-3ex}}
        \centering
        \includegraphics[width=\columnwidth]{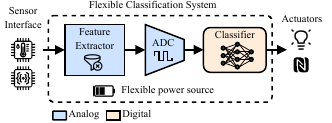}\vspace{-2ex}
        \label{fig:system_analog}
    \end{subfigure}
    \caption{Abstract overview of a flexible classification system, with feature extractors in the (a) digital, or (b) analog domain}
    \label{fig:system}
    \vspace{-3ex}
\end{figure}

Within these constraints, a typical \gls{fe}-based healthcare monitoring system integrates flexible sensors for physiological signal acquisition, quantization via \glspl{adc}, and on-device feature extraction and \gls{ml} classification. 
Typically, such systems are mostly implemented in digital logic (Fig.~\ref{fig:system_digital}).
Although each component plays a critical role, most existing \gls{fe} approaches concentrate solely on the design and optimization of either the sensor interface or the \gls{ml} classifier~\cite{Alkhalil:BioCAS:2022:FlexibleSAR,Wang2017FlexibleSE,Anzanpour2017SelfawarenessIR,Yang:RSC2019:flexmonitoring,yoon:nature2016:flexibleSensorPatch,chen:IEEE2020:FlexibleSensors, Ozer:2019:Bespoke, Iordanou:Nature2024:PRAGMATIC:evolutionary_tiny_classifiers,Armeniakos:TC2023:codesign,Kokkinis:TC:2024:enabling,Afentaki:ICCAD23:hollistic,Balaskas:ISQED2022:axDT,Kokkinis:DATE2023}. 
Acknowledging this limitation,~\cite{duarte:ASPDAC2024:pruneBinaries,Afentaki:2024LES:ReducingAF,Armeniakos:DATE:2024:sensor,Mrazerk:ICCAD2024} co-design the analog interface with the classifier, achieving notable savings in both the interface and the overall system.
However, the feature extractor, which plays a vital role in increasing classification performance and maintaining acceptable hardware requirements, is yet overlooked by the state of the art.
In fact, as we demonstrate in~\cite{afentaki2025islped}, feature extractors form the area bottleneck in such flexible systems (\cref{fig:system_digital}) and can incur prohibitive area and power costs, potentially challenging the feasibility of the entire system.

In our work, we address these challenges with an automated feature-to-classifier co-design framework for ultra-area-efficient mixed-signal flexible healthcare systems.
Our approach jointly optimizes all core components---including \glspl{adc}, feature extractors, and classifier---aiming to reduce the hardware overheads of \gls{adc} and feature extraction.
First, to mitigate its elevated area overhead,
we implement feature extraction directly in the analog domain--as shown in \cref{fig:system_analog}--and, to the best of our knowledge, design the first analog feature extractors in \gls{flexic} technology.
Next, we design a Successive Approximation Register (SAR) \gls{adc} optimized for our system, where all analog features are quantized and stored in buffers for processing by the classifier.
For classification, we employ shallow digital \glspl{mlp}\blue{---tailored to the application, i.e., bespoke designs---}which can
deliver top accuracy within realistic hardware overheads~\cite{afentaki2025islped}.
Finally, we propose a hardware-aware feature selection technique embedded within \gls{mlp} training, which minimizes feature extraction cost while maximizing accuracy.
\textbf{Our novel contributions within this work are as follows}:
\begin{enumerate}[topsep=0pt,leftmargin=*]
    \item 
    We design, for the first time, analog feature extractors in \gls{flexic} targeting conformable classification systems.
    \item 
    We propose the first \blue{holistic feature-to-classifier co-design and co-optimization framework} for mixed-signal flexible systems, where feature extraction is implemented in the analog and classification in the digital domain, to balance cost and accuracy,
    along with a novel hardware-aware feature selection embedded within \gls{mlp} training.
    \item 
    We evaluate our work on relevant healthcare datasets and demonstrate that our co-design framework enables flexible systems with high accuracy and ultra-low area and energy-per-inference requirements, making them ideal for low-cost, conformal, and disposable healthcare wearables.
\end{enumerate}

\section{Background on FlexICs}
\label{sec:background_related}
\label{sec:flexible_electronics}

\begin{figure}[t]
    \centering
    \includegraphics[width=0.4\linewidth]{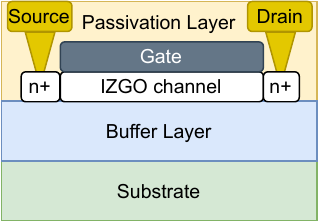}
    \caption{Process schematic illustrating photolithographic patterning on IGZO semiconductors and metallic interconnects.}
    \label{fig:FE_background}
    \vspace{-3ex}
\end{figure}

Pragmatic's \gls{flexic} technology enables the fabrication of mechanically flexible circuits on ultra-thin polyimide substrates using \gls{igzo} \glspl{tft}~\cite{tahoori2025computing, ozer:nature2024:bendableRiscV,flexic_gen3}.
\gls{igzo} \glspl{tft} are manufactured using low-temperature photolithography and cost-effective equipment, eliminating the need for rigid silicon wafers, high-temperature processes, and protective packaging, as illustrated in Fig~\ref{fig:FE_background}.
This streamlined process significantly reduces environmental impact, lowers production costs, and shortens fabrication time---from over 30 weeks in traditional silicon technology to just a few days for \glspl{flexic}~\cite{ozer:nature2024:bendableRiscV}.
The produced circuits are mechanically robust, bendable to a radius as small as \SI{3}{\milli\meter}, and well-suited for conformal, disposable electronics in healthcare applications.

Despite these advantages, \gls{igzo}-based \glspl{flexic} face key limitations compared to CMOS technology.
Their relatively large feature sizes ($600\text{–}800$\si{\nano\meter}) and low integration density lead to increased power consumption with tight area requirements~\cite{tahoori2025computing,ozer:nature2024:bendableRiscV, ARM:2021:PlasticARM}.
Due to the absence of stable p-type transistors, \gls{fe} are restricted to n-type devices and unipolar logic.
As a result, resistive-load NMOS logic incurs high--especially static--power consumption and high circuit latencies, leading to elevated energy demands as well.
As a result, current \gls{flexic} designs are limited to moderate complexity---typically only a few thousand gates---and require careful hardware-software co-design to simplify logic, reduce gate count, and limit memory elements, which remain costly in \gls{fe}~\cite{tahoori2025computing,Ozer:2019:Bespoke}. 
Such constraints highlight the need for specialized, lightweight design methodologies---especially in the healthcare domain where area, energy autonomy, and flexibility are critical.

\section{Motivation}
\label{sec:motivation}
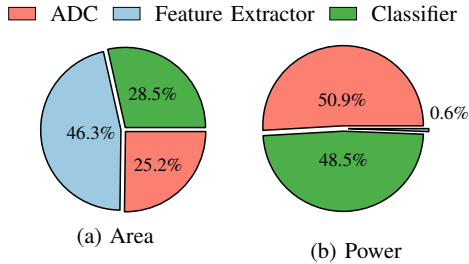
\begin{figure}[t!]
    \centering
    \resizebox{.75\columnwidth}{!}{
        \begin{tikzpicture}

\node at (0, 0) {
    \centering
    \begin{tikzpicture}
        \matrix[column sep=1ex] {
            \node[draw=black, fill=areaADC, minimum width=0.5cm, minimum height=0.2cm] {}; &
            \node[font=\large] {ADC}; &

            \node[draw=black, fill=areaFEx, minimum width=0.5cm, minimum height=0.2cm] {}; &
            \node[font=\large] {Feature Extractor}; &

            \node[draw=black, fill=areaClf, minimum width=0.5cm, minimum height=0.2cm] {}; &
            \node[font=\large] {Classifier}; \\
        };
    \end{tikzpicture}
};

\node at (0, -2.5) { 
  \begin{tikzpicture}
    \begin{scope}[xscale=1]
      \begin{scope}[shift={(-2, 0)}] 
        \node {
            \begin{subfigure}{0.4\columnwidth}
                \centering
                \begin{tikzpicture}
                    \pie[
                        radius=1.5,
                        explode=0.05,
                        color={areaClf, areaFEx, areaADC}
                    ] 
                    {
                        28.5/,
                        46.3/,
                        25.2/
                    } 
                \end{tikzpicture}
                \captionsetup{font=large}
                \caption{Area}
            \end{subfigure}
        };
      \end{scope}

      \begin{scope}[shift={(2.5, 0)}] 
        \node {
            \begin{subfigure}{0.4\columnwidth}
                \centering
                \begin{tikzpicture}
                    \pie[
                        radius=1.5,
                        explode=0.05,
                        color={areaADC,areaClf}
                    ] 
                    {
                        50.9/,
                        48.5/
                    }
                    \pie[
                        hide number,
                        radius=1.5,
                        change direction,
                        explode=0.1, 
                        color={areaFEx}
                    ]
                    {
                        0.6/
                    }
                    \node[
                        anchor=west,
                    ] at (1.5, 0.3) {0.6\%};
                \end{tikzpicture}
                \captionsetup{font=large}
                \caption{Power}
            \end{subfigure}
        };
      \end{scope}
    \end{scope}
  \end{tikzpicture}
};

\end{tikzpicture}
    }
    \caption{Area and power breakdown of an \acrshort{mlp}-based flexible system on the WESAD dataset~\cite{Schmidt2018IntroducingWA:wesad}.
    We observe that the digital feature extractors consume a major portion of the total area, whereas the \acrshortpl{adc} constitute the system's power bottleneck.}
\label{fig:motiv-piechart}
\end{figure}


Feature selection is often treated purely from an algorithmic perspective, without consideration for its hardware implications~\cite{Shiyi:IoT2022:StressMonitoring}, leading to designs that, in \gls{fe}, may be infeasible for the underlying area and power constraints.
Indicatively, we present an example in \cref{fig:motiv-piechart} using the WESAD stress-monitoring dataset~\cite{Schmidt2018IntroducingWA:wesad}, where we perform feature selection with the statistical Fisher score algorithm, similar to~\cite{Shiyi:IoT2022:StressMonitoring}.
We then train a \gls{mlp} classifier on the extracted features and design the digital circuits that implement the required feature extractors and the resulting \gls{mlp}.
As shown in \cref{fig:motiv-piechart}, the feature extraction consumes the central portion of the total area, at $46\%$, which rises to $74\%$ when accounting for \gls{adc} costs.
In addition, \glspl{adc} account for $51\%$ of the total power consumption of the system.
\emph{We therefore conclude that focusing solely on the \gls{ml} classifier is suboptimal and can challenge the design feasibility of such flexible systems.
To achieve the utmost hardware efficiency--as required in \gls{fe} to even enable feasibility--a holistic feature-to-classifier optimization is mandatory.}


\section{\blue{Flexible Mixed-Signal System Architecture}}
\label{sec:system}
\begin{figure}[t]
    \centering
    \includegraphics[width=0.75\columnwidth]{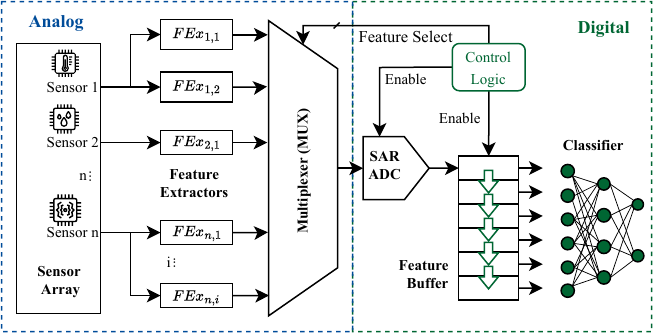}
    \caption{Overview of our proposed flexible mixed-signal system architecture}
    \label{fig:system_circuit}
\end{figure}

An abstract overview of our proposed system architecture is illustrated in \cref{fig:system_circuit}.
Assuming an array of biosensors, the analog feature extraction circuits process the incoming sensor signals and compute the required statistics (features) over a predefined window size.
Each feature for each sensor is computed on dedicated hardware, with all features computed concurrently.
At the end of the timing window, the analog features (outputs of the analog feature extraction circuits) are multiplexed through the SAR \gls{adc}, and their quantized values are stored in a buffer for subsequent processing by the digital \gls{mlp} classifier.
A small digital control logic block (e.g., a few counters, a decoder) orchestrates the entire operation.

\subsection{Analog Feature Extractors}
\label{sec:feature_extractor}

\begin{figure}
    \centering
    \resizebox{0.8\columnwidth}{!}{
        \begin{tikzpicture}
\begin{axis}[
    ybar,
    ymin=-1,
    ymax=20,
    bar width=7pt,
    enlarge x limits=0.05,
    ylabel={\shortstack{Relative\\Frequency (\%)}},
    symbolic x coords={Min, Max, Sum, Mean, Range, SumPeaks, MuPeaks, NPeaks, Std. Dev., Skewness, Kurtosis, Median},
    xtick=data,
    xticklabel style={font=\normalsize, rotate=45, anchor=east},
    yticklabel style={font=\normalsize},
    nodes near coords align={vertical},
    width=\linewidth,
    height=4cm,
    ymajorgrids=true,
    major grid style={dashed, gray!40},
    tick style={/pgfplots/major tick length=4pt},
    tick align=inside
]

\addplot+[draw=black, fill=blue!50] coordinates {
    (Min, 17.19)
    (Max, 16.68)
    (Sum, 13.58)
    (Mean, 12.99)
    (Range, 9.38)
    (SumPeaks, 8.51)
    (MuPeaks, 7.86)
    (NPeaks, 7.64)
    (Std. Dev., 4.31)
    (Skewness, 1.04)
    (Kurtosis, 0.82)
    (Median, 0.00)
};

\end{axis}
\end{tikzpicture}
    }
    \caption{Most frequent features selected in the accuracy–area Pareto-optimal flexible classifiers from~\cite{afentaki2025islped} across various healthcare datasets}
    \label{fig:feature_histogram}
\end{figure}
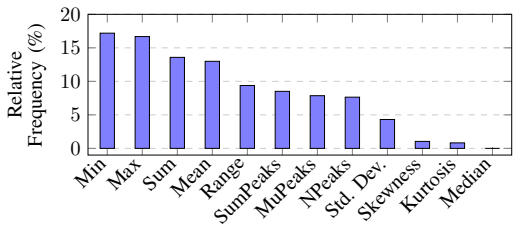

In~\cite{afentaki2025islped}, we conducted a comprehensive exploratory study on the implications and hardware overheads of feature extraction in the design of \gls{ml}-based flexible healthcare wearable systems. \cref{fig:feature_histogram} shows the frequency of the most commonly selected features in the accuracy–area Pareto-optimal designs identified in~\cite{afentaki2025islped}. As observed, simple features such as maximum, minimum, mean, and sum appear more frequently in Pareto-optimal solutions and are always present in the most accurate ones. Therefore, in our work, we design the analog equivalents of these four statistical functions in FlexIC technology, and integrate them into our co-design process.

\subsubsection{Maximum (Max)}
The Max block uses a peak-detector to capture and hold the highest value of the input signal. As shown in Fig.~\ref{fig:Statistical_Features}(a), it comprises a diode and a hold capacitor. When \(v_{\text{in}} > v_C + V_{TH}\), the diode conducts and charges the capacitor to the new peak; when \(v_{\text{in}} < v_C\), the diode is reverse-biased, blocking current and preserving the stored peak.

Since the FlexIC PDK provides only n-type transistors and no discrete diodes, we implement the diode function with a diode-connected n-type transistor (gate tied to drain). This introduces a headroom requirement: conduction requires \(V_{GS}\!\ge\!V_{TH}\), so any path through this element incurs an effective forward drop of roughly \(V_{TH}\), which reduces available signal swing and sets a minimum rectifiable amplitude. In addition, \emph{leakage}, dominated by subthreshold conduction when \(V_{GS}\!<\!V_{TH}\) causes residual discharge that degrades low-level accuracy; the droop rate is approximately \(\dot v_C \approx -I_{\text{leak}}/C\). We mitigate these effects by using longer-channel devices.

Capacitor sizing trades hold time against tracking speed: larger \(C\) increases the hold time (lower \(|\dot v_C|\)) but slows response to rapid input changes, whereas smaller \(C\) enables faster tracking at the cost of increased droop and reduced hold time.
A windowing switch is included to reset the function. It isolates the hold capacitor at each window boundary and clears the previous state before the next window. For the Max configuration, the reset discharges the hold capacitor to the low reference \(V_l\), after which it can charge within the window to track the maximum.

\subsubsection{Minimum (Min)}
\begin{figure}[t]
    \centering
    \includegraphics[width=0.9\columnwidth]{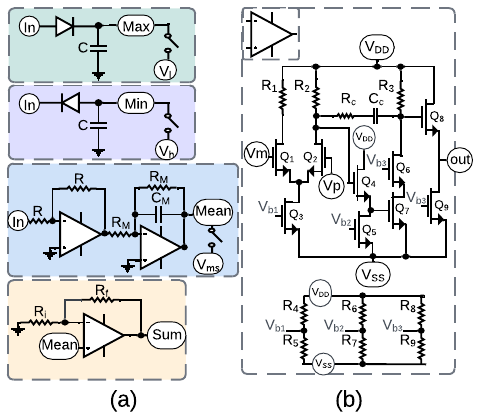}
    \caption{(a) Analog feature circuits of Max, Min, Mean, and Sum. (b) Op-amp implementation in the FlexIC PDK}
    \label{fig:Statistical_Features}\vspace{-2ex}
\end{figure}

The Min block is a valley detector that captures and holds the lowest value of the input over each observation period. As in Fig.~\ref{fig:Statistical_Features}(a), the diode–capacitor network is oriented to pull the capacitor node down when the input decreases. When \(v_{\text{in}} < v_{C} - V_{TH}\), the diode-connected n-type transistor conducts and discharges the capacitor to the new minimum; when \(v_{\text{in}} > v_{C}\), the device is reverse-biased, blocking current and preserving the stored minimum.

Similar to the Max implementation, the diode is implemented with a diode-connected n-type transistor. This introduces a headroom condition: updates occur only when the input drop exceeds \(V_{TH}\), limiting sensitivity to small dips.
In addition, leakage causes the stored value to drift from the true minimum over time, at a rate approximately \(|\dot v_{C}|\!\approx\! I_{\text{leak}}/C\).

A windowing switch is included, as in the Max design. In the Min configuration, however, the reset pre-charges the hold capacitor to the upper reference \(V_h\), after which it can only discharge within the window to track the minimum.

\subsubsection{Mean} The mean calculation circuit (Fig.~\ref{fig:Statistical_Features}(a)) is based on an op-amp integrator, which continuously accumulates the input signal, summing values over a defined period according to the integration equation.
This accumulated output approximates the mean when normalized by the integration period $T$ or scaled by $\frac{1}{R_MC_M}$, analogous to dividing a discrete sum by the number of samples.
An inverter adjusts the polarity of the input signal, which is then inverted again by the integrator, ensuring the final output matches the expected mean value.
\begin{equation}
\text{Mean } (\mu) = \frac{1}{N} \sum_{i=1}^{N} x_i \approx \frac{1}{R_MC_M} \int_0^T x(t) \, dt,
\end{equation}
where \( x_i \) represents discrete input samples, \( N \) is the sample count, \( x(t) \) is the continuous-time input signal, \( T \) is the integration period, and \( R_M \) and \( C_M \) are the resistor and capacitor values in the integrator circuit, respectively.

The windowing switch also performs a reset at each window boundary. For the Mean block, we reset to \(V_{ms}\), the mid-scale (virtual-ground) level of the op-amp output range (typically \(V_{ms} \!\approx\! (V_{\mathrm{DD}}+V_{\mathrm{SS}})/2\)), so each window starts centered.

\subsubsection{Sum}
The sum operation is similar to the mean calculation but scaled by the number of samples $N$.
Leveraging this, we scale the mean output to approximate the sum.
However, since analog circuits are limited by the op-amp's output swing, the scaling factor is chosen to keep the resulting output within the op-amp's acceptable operating range.
\begin{equation}
\text{Sum} = \mu \times N, \quad N = 1+\frac{R_f}{R_i}.
\label{sum}
\end{equation}
The scaling is implemented using a non-inverting amplifier configuration, as shown in Fig~\ref{fig:Statistical_Features}(a), where the gain follows the relationship in~\eqref{sum}.
By selecting appropriate resistor values $R_f$ and $R_i$, the circuit scales the mean output to produce the sum.
Since the Sum is derived directly from the Mean output, it does not require a separate reset for each window.

\subsubsection{op-amp}
Designing the op-amp in FlexIC presents challenges, as the technology includes only n-type transistors and lacks p-type devices. Despite this limitation, a two-stage op-amp with an additional buffer stage is implemented, as shown in Fig.~\ref{fig:Statistical_Features}(b).
The op-amp consists of a differential amplifier stage, which includes a pair of n-type transistors forming the differential pair \((Q_1\),\(Q_2\)), along with a tail-bias transistor \(Q_{3}\) and passive resistive loads \(R_{1}\) and \(R_{2}\). 
A common-source amplifier follows this stage to enhance the gain of the op-amp.
In this configuration, two transistors are used: one to stabilize the operating point and set the current of the common-source amplifier \(Q_6\), and the other to serve as the input transistor \(Q_7\), receiving the signal from the preceding stage.
To ensure proper operation of \(Q_7\), we add a level-shifting stage using transistors \(Q_4\) and \(Q_5\) to set the node at the required DC bias level.
The output of the differential amplifier is coupled to the input of the common-source stage through a coupling capacitor \(C_c\) and resistor \(R_c\), which help maintain stability and improve phase margin.
To ensure low output impedance, a buffer stage is included with two transistors \(Q_8\) and \(Q_9\).
It isolates the op-amp’s high-gain stages from the load, reducing the risk of loading effects and improving driving capability.

\subsection{Flexible SAR ADC}
\label{sec:adc}

Various ADC designs have been explored in FE, including Flash~\cite{JAMSHIDIROUDBARI:2010:FlashFlexible}, Sigma-Delta~\cite{Garripoli:2017:DeltaSigmaADC}, Binary Search~\cite{duarte:ASPDAC2024:pruneBinaries}, and SAR~\cite{Alkhalil:BioCAS:2022:FlexibleSAR}.
SAR ADC is attractive due to its moderate hardware complexity, scalability to different resolutions, and low static power dissipation.
Therefore, we design a custom SAR ADC fully tailored to our application requirements (e.g., speed, precision, input range), thereby maximizing hardware efficiency in our systems.

\begin{figure}[t]
    \centering
    \includegraphics[width=.75\columnwidth]{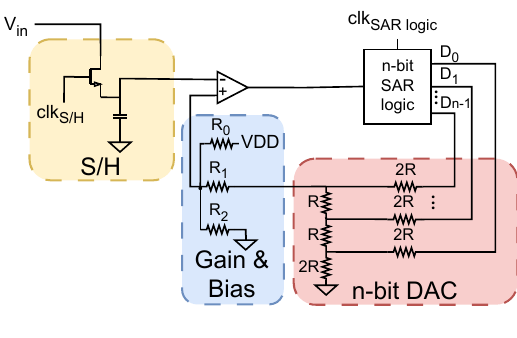}
    \caption{Schematic of an $n$-bit SAR ADC}
    \label{fig:SAR}\vspace{-2ex}
\end{figure}

A SAR ADC determines the $n$-bit digital output code in $n$ cycles. In each cycle, the SAR logic sets the current bit and drives a digital-to-analog converter (DAC).
The DAC output is compared to the sampled analog input by a comparator, and the SAR logic updates the code accordingly.
If the comparator output is high, the current bit remains set; otherwise, it is cleared before moving to the next bit.
Once the LSB has been processed, the resulting SAR logic value is the ADC output.

An $n$-bit schematic of our designed SAR ADC is shown in \cref{fig:SAR}. 
The control logic is implemented in Verilog RTL as two $n$-bit registers and mapped to PragmatIC’s FlexIC standard digital cell library.
This cell library is based on a resistive-load logic architecture, which uses a fixed resistor for the pull-up.
The DAC is implemented as an $R$–$2R$ ladder network, producing $2^n$ discrete voltage levels.
A subsequent gain-and-bias stage scales and offsets the DAC output to align with the output range of the analog feature circuits.
Although our ADC can operate significantly faster, we observe that across all healthcare datasets examined in Section~\ref{sec:evaluation}, a conversion time of $0.5$ms is sufficient for real-time monitoring. 

\subsection{Flexible Digital MLP Classifier}
\label{sec:classifier}
\begin{figure}[t]
    \centering
    \includegraphics[width=0.75\columnwidth]{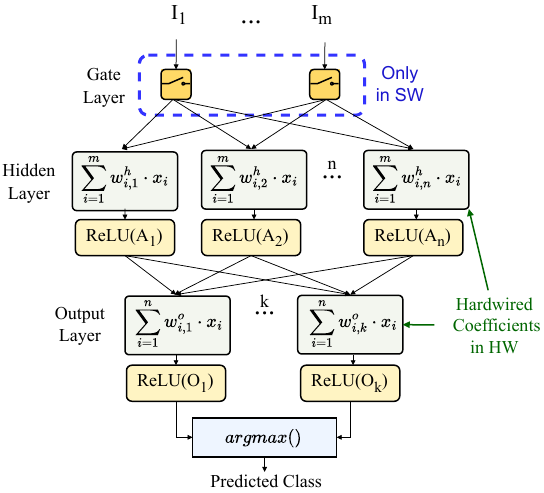}
    \caption{Bespoke fully-parallel \acrshort{mlp} design}
    \label{fig:mlp}\vspace{-2ex}
\end{figure}

The digital part of our system involves an \gls{mlp} classifier (see \cref{fig:system}b), due to its effectiveness in delivering high accuracy at relevant applications~\cite{Kokkinis:TC:2024:enabling,Armeniakos:TC2023:codesign, afentaki2025islped}.
Leveraging the ultra-low manufacturing and non-recurring engineering (NRE) costs of flexible electronics (FE), we implement our digital \gls{mlp} as a fully-parallel bespoke architecture~\cite{Ozer:2019:Bespoke,Ozer:Nature:2020,Kokkinis:TC:2024:enabling}, as shown in \cref{fig:mlp}.
Bespoke \gls{ml} circuits hardwire the trained coefficients into the circuit, enabling significant area savings compared to conventional designs~\cite{Armeniakos:DATE2022:axml}, and facilitating further logic simplification through constant propagation during synthesis.
Fully-parallel \glspl{mlp} instantiate one hardware neuron per software neuron, with each containing one multiplier per trained weight, followed by a precision-optimized adder tree for product accumulation.
All neurons operate concurrently, eliminating the need for costly memory elements in \gls{fe}.
Such architectures are favorable to unstructured pruning, with direct hardware savings due to the lack of folding.
We later demonstrate how unstructured pruning-aware retraining is embedded into our framework for further area reductions.

\section{Feature-to-Classifier Co-Design Framework}
\label{sec:framework}
\begin{figure}
    \centering
    \includegraphics[width=\columnwidth]{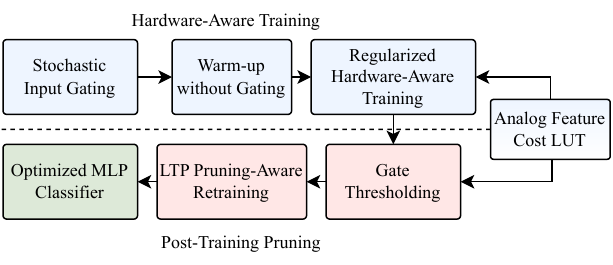}
    \caption{Algorithmic overview of our proposed hardware-aware co-design framework}
    \label{fig:framework_flow}\vspace{-2ex}
\end{figure}

In this section, we present our automated co-design framework for training the flexible \gls{mlp} classifier.
An algorithmic overview of our framework is presented in \cref{fig:framework_flow}. 
It first introduces a novel hardware-aware feature selection technique embedded within training (\cref{sec:framework_training_technique}), aiming to reduce the analog feature extractor costs and retain high classification accuracy within a unified differentiable step.
Then, leveraging our system's architecture, pruning with retraining steps follow (\cref{sec:lottery_ticket}) to further maximize area efficiency.

\subsection{Differentiable Feature Selection \& Training}
\label{sec:framework_training_technique}
Our feature selection mechanism is inspired by \gls{nas} methods~\cite{darts}, which typically focus on layer or block selection.
Adapted for our purposes,
our approach operates at the feature level, embedding a differentiable stochastic gating layer into the input stage of an \gls{mlp}, as shown in \cref{fig:mlp}. 
By integrating cost-aware regularization into the gating mechanism, the technique enables end-to-end optimization of both accuracy and analog feature extraction cost within the training process.

\subsubsection{Stochastic Feature Gating}
\label{subsec:gates}
Assuming all input features for all sensors $\mathbf{x} \in \mathbb{R}^d$, we introduce a trainable gating vector $\mathbf{z} \in [0, 1]^d$ applied as a multiplicative factor to the input: 
\begin{equation}
\tilde{\mathbf{x}} = \mathbf{z} \odot \mathbf{x}
\end{equation}

Each gate $z_i$ represents the inclusion probability of feature $x_i$ and is modeled as a stochastic binary variable.
Features with lower gate values (i.e., close to zero) contribute negligibly during training and can therefore be considered non-important.
To enable gradient-based optimization, this binary behavior is approximated using the Concrete (Gumbel–Sigmoid) distribution~\cite{gumbel_sigmoid}.
During training, gates are stochastically sampled as:
\begin{equation}
z_i = \text{clip}(s_i, 0, 1),
\end{equation}
\begin{equation}
\label{eq:sample_sigma}
s_i = \frac{1}{1 + \exp\left( -\frac{ \log u_i - \log(1 - u_i) + \log \alpha_i }{\gamma} \right)},
\end{equation}
where $\log \alpha_i$ is the trainable parameter (i.e., logit) controlling the openness of the gate,  
$u_i \sim \mathcal{U}(0, 1)$ is a uniform random variable,  
and $\gamma$ is a hyperparameter controlling the relaxation.  
As $\gamma$ approaches $0$, the distribution becomes more discrete, yielding gate values closer to ${0,1}$ while still maintaining differentiability for optimization.
By progressively lowering $\gamma$, we can gradually separate essential features from insignificant ones based on their corresponding gate values.

After training has converged, the gates provide interpretable importance scores per feature.
Specifically, the stochastic sampling above is replaced by simple deterministic gating:  
\begin{equation}
z_i = \sigma(\log \alpha_i),
\end{equation}
where $\sigma(\cdot)$ is the logistic function mapping logits to probabilities in $(0,1)$.
Thus, during inference the gating vector $\mathbf{z}$ is fixed and modulates the contribution of each feature.

The gating layer is placed directly between the input and the first hidden layer (see \cref{fig:mlp}), ensuring that all subsequent computations operate only on the modulated input $\tilde{\mathbf{x}}$, thereby minimally impacting the network’s architecture.

\subsubsection{Area-Informed Regularized Training}
\label{subsec:regularization}
In order to introduce hardware-awareness into the training and embed the feature extraction cost into the stochastic gates, we employ regularization by adding a cost-aware term to our loss function.
Specifically, we collect the area of our analog feature extraction circuits (see \cref{sec:feature_extractor}) in a \gls{lut}, which is then proportionally added to the training loss.  

Let $\mathbf{c} \in \mathbb{R}^d$ be the vector of per-feature area costs.  
Accounting for the gated contribution of each feature, the expected cost $\mathcal{L}_{\text{cost}}$ is combined with the task loss as follows:  
\begin{equation}
\label{eq:regularization}
\mathcal{L} = \mathcal{L}_{\mathrm{CE}} + \lambda \cdot \mathcal{L}_{\text{cost}}, \; \;
\mathcal{L}_{\text{cost}} = \sum_{i=1}^{d} \sigma(\log \alpha_i) \cdot c_i,
\end{equation}
where $\mathcal{L}_{\mathrm{CE}}$ is the cross-entropy loss,  
and $\lambda$ a scalar hyperparameter controlling the accuracy-cost trade-off.  
By directly embedding feature extraction costs into the optimization objective, the model is incentivized to discover feature subsets that are jointly optimal for accuracy and hardware efficiency.  

Since regularization is known for destabilizing training~\cite{wen2016learning}, we employ a warm-up phase of $k$ epochs during which gradients through the gate layer are detached, keeping them fixed at their initialized values:  
\begin{equation}
z_i = \text{stop\_gradient}(z_i) \quad \text{if epoch} < k.
\end{equation}
This allows the \gls{mlp} weights to adapt to the objective loss ($\mathcal{L}_{\mathrm{CE}}$) before stochastic gating--and therefore input destabilization--is introduced.

\subsubsection{Feature Pruning via Gate Removal}
\label{subsec:gate_removal}
After convergence, we perform feature selection by removing gates whose value is below a threshold $\tau$:
\begin{equation}
\label{eq:gate_pruning}
\hat{z}_i = 
\begin{cases}
1 & \text{if } z_i > \tau \\
0 & \text{otherwise}.
\end{cases}
\end{equation}
This thresholding produces a sparse and hardware-efficient input representation,  
preserving high-importance features while eliminating those with low contribution to accuracy.
By progressively iterating over various thresholds $\tau\!\in\![0, 1]$, we obtain a Pareto-front of networks that trade off accuracy and feature extraction cost in a fully automated, hardware-aware manner.

\subsection{Lottery-Ticket Pruning-Aware Retraining}
\label{sec:lottery_ticket}
Following feature selection and training, where the analog frontend costs are optimized, we aim to reduce the significant contribution of the digital classifier to the total system area.
Due to its bespoke, fully parallel architecture, unstructured pruning can be directly exploited in this context to reduce the number of parameters, thereby yielding high hardware gains by eliminating multipliers and minimizing the area of accumulators~\cite{Kokkinis:TC:2024:enabling}.
To that end, we adopt pruning-aware retraining in the form of \gls{ltp}, inspired by the Lottery Ticket Hypothesis~\cite{frankle2018lottery},
which stipulates that within a dense, randomly-initialized neural network, there exists a sparse subnetwork capable of matching the original performance when trained in isolation.

Let $\mathbf{W} \in \mathbb{R}^n$ denote the vectorized parameters of the trained \gls{mlp}, and let $\mathbf{m} \in \{0,1\}^n$ be a binary pruning mask, where $m_j = 0$ indicates removal of weight $W_j$.
We perform iterative magnitude pruning by updating $\mathbf{m}$ according to:
\begin{equation}
m_j =
\begin{cases}
0 & \text{if } |W_j| \leq \kappa_t \\
1 & \text{otherwise},
\end{cases}
\end{equation}
where $\kappa_t$ is the pruning threshold at iteration $t$, chosen to achieve a target sparsity $s_t \in [0,1]$.
After each pruning step, the remaining weights are reset to their pre-training values $\mathbf{W}_0 \odot \mathbf{m}$, and the sparse subnetwork is retrained for a reduced number of epochs, progressively increasing $s_t$ until the desired final sparsity is reached.

\section{Results \& Analysis}
\label{sec:evaluation}

\subsection{Experimental Setup}
\label{sec:experimental_setup}
We evaluate our proposed framework on 3 popular healthcare benchmarks related to the use-case of stress-monitoring,
where physiological data are extracted from biosensors to infer the stress levels of individuals.
Specifically, we use the WESAD dataset~\cite{Schmidt2018IntroducingWA:wesad}--the most popular and common dataset for stress-monitoring applications--alongside Stress-In-Nurses~\cite{hosseini:Nature2022:stress_in_nurses}, and Stress~Predict dataset (or SPD)~\cite{spd:stressdataset}.
Though, our approach can be seamlessly extended to any relevant far-edge and/or healthcare application and dataset.

Input data are in floating-point format and normalized within the $[0, 1]$ range.
Feature extraction is simulated at high level in Python using a non-overlapping sliding window of \SI{1}{\second}.
K-fold cross-validation is used on all considered datasets, using $80\%$ of subjects for training, and the rest $20\%$ for test on unseen individuals' data.
The accuracy is reported on the test set hereafter.
For training, we use Tensorflow and consider
an \gls{mlp} with one hidden layer of $100$ neurons, such that high enough computational accuracy can be achieved, while any redundancy from the large neuron count can be removed by pruning-aware retraining.
The Adam optimizer with a learning rate of $0.001$ is used for $50$ epochs of training ($10$ for the pruning-aware retraining), along with early-stopping.
The regularization parameter $\lambda$ (see \eqref{eq:regularization}) and $\gamma$ (see \eqref{eq:sample_sigma}) undergo hyperparameter tuning via Bayesian optimization within empirical ranges.
Finally, the gate pruning thresholds ($\tau$ in \eqref{eq:gate_pruning}) are swept within $\{0.01, 0.05, 0.1, 0.2, 0.5\}$.
\gls{mlp} weights are quantized post-training to $8$ bits, while input features are quantized to $4$ bits, matching the precision that we use for the ADC.
Such low input/ADC precision is common in FE applications~\cite{ozer2023malodour,Kokkinis:TC:2024:enabling,afentaki2025islped} and is sufficient to maintain high accuracy while incurring low analog interfacing costs.




\begin{table}[t] 
\centering
\caption{Dimensions of op-amp components}
\label{tab:opamp_sizing}
\setlength{\tabcolsep}{6pt}
\begin{tabular}{l l}
\toprule
\multicolumn{1}{c}{\textbf{Component}} & \multicolumn{1}{c}{\textbf{Size}} \\
\midrule
$Q_1$ to $Q_7$            & W = 20\,\si{\micro\meter}, L = 1.2\,\si{\micro\meter} \\
$Q_8$                     & W = 200\,\si{\micro\meter}, L = 0.6\,\si{\micro\meter} \\
$Q_9$                     & W = 1\,\si{\micro\meter}, L = 1.2\,\si{\micro\meter} \\
$R_1$, $R_2$ = 368.5\,\si{\kilo\ohm}  & W = 2.1\,\si{\micro\meter}, L = 4.4\,\si{\micro\meter} \\
$R_3$ = 837.6\,\si{\kilo\ohm}         & W = 2.1\,\si{\micro\meter}, L = 10\,\si{\micro\meter} \\
$R_4$ = 2.3\,\si{\mega\ohm}           & W = 1\,\si{\micro\meter}, L = 12.3\,\si{\micro\meter} \\
$R_5$ = 1\,\si{\mega\ohm}             & W = 1\,\si{\micro\meter}, L = 5.3\,\si{\micro\meter} \\
$R_6$ = 15.4\,\si{\mega\ohm}          & W = 0.6\,\si{\micro\meter}, L = 48.2\,\si{\micro\meter} \\
$R_7$ = 10\,\si{\mega\ohm}            & W = 0.6\,\si{\micro\meter}, L = 28.5\,\si{\micro\meter} \\
$R_8$, $R_9$ = 20\,\si{\mega\ohm}     & W = 0.6\,\si{\micro\meter}, L = 61.4\,\si{\micro\meter} \\
$R_C$ = 50\,\si{\kilo\ohm}            & W = 50\,\si{\micro\meter}, L = 15\,\si{\micro\meter} \\
$C_C$ = 0.95\,\si{\pico\farad}                 & W = 14\,\si{\micro\meter}, L = 15\,\si{\micro\meter} \\
\bottomrule
\end{tabular}
\end{table}





\begin{table}
\caption{Design properties of the SAR ADC}
\label{tab:ADC_sizing}
\centering
\setlength{\tabcolsep}{4pt}
\begin{tabular}{l l l}
\toprule
\textbf{Component} & \textbf{Element} & \textbf{Size} \\
\midrule
\textbf{S/H} & Transistor & w = 15\si{\micro\meter}, l = 600\,nm \\
             & Capacitor  & cap = 2\,pF \\
\midrule
\textbf{DAC} & Resistor & R = 1.5\,M$\Omega$ \\
                 & Resistor & $R_0$, $R_2$ = 6\,M$\Omega$ \\
                 & Resistor & $R_1$ = 4.5\,M$\Omega$ \\
\midrule
\textbf{Comparator} & \multicolumn{2}{l}{We use the op-amp presented in Table~\ref{tab:opamp_sizing}} \\
\midrule
\textbf{SAR Digital Logic} & \multicolumn{2}{l}{Std-cell based design with Gen-3 FlexIC} \\
\bottomrule
\end{tabular}\vspace{-2ex}
\end{table}

For analog and mixed-signal simulations, we use the Cadence Spectre simulator with the Gen3 PragmatIC FlexIC 1.0.0 PDK~\cite{flexic_gen3}.
The supply voltage is set to \SI{3}{\volt}. 
For digital synthesis, timing, and power simulations (e.g., for the \gls{mlp}), Synopsys Design Compiler, VCS, and PrimeTime are used.
Synthesized designs are mapped to PragmatIC’s Gen-3 FlexIC PDK characterized at \SI{3}{\volt}~\cite{flexic_gen3}.
Our classifiers are synthesized at a relaxed clock period using the \texttt{compile\_ultra} command and target area optimization.
A base clock of \SI{10}{\kilo\hertz} is considered, and our systems target real-time performance, i.e., producing a stress prediction every $1$ s.
Specifically, after the last sample of the window is received, our system completes the classification (feature-to-MLP) within less than $20$ ms.
Power-gating is used for idle components. 


The area of the analog circuits is estimated from the pre-layout device dimensions ($W$–$L$) of the components.
The op-amp sizing is listed in Table~\ref{tab:opamp_sizing}, \blue{\(Q_8\) is intentionally wider in the buffer stage to increase the output-swing range.}
The opamp's power consumption is measured in a closed-loop configuration.
The ADC is designed for $4$-bit quantization, with transistor and resistor dimensions provided in Table~\ref{tab:ADC_sizing}.
Although $R_0$, $R_1$, and $R_2$ should be theoretically equal to keep the signal within $1$–$2$V, the inherent circuit attenuation requires adjusting these values.
To compensate, we set $R_0$ and $R_2$ larger than $R_1$.
By this adjustment, the DAC output ranges from $0.98$V to $1.95$V.



\subsection{Analog Components Evaluation}
\label{sec:results}



\begin{table}[t]
\centering
\caption{Op-amp performance metrics}
\label{tab:opamp_results}
\setlength{\tabcolsep}{4pt}
\begin{tabular}{l l}
\toprule
\multicolumn{1}{c}{\textbf{Metric}} & \multicolumn{1}{c}{\textbf{Value}} \\
\midrule
Gain (after buffer stage)        & 28.6\,dB \\
Unity Gain Bandwidth              & 902\,kHz \\
Phase Margin                      & 58\textdegree \\
Slew Rate (Rising / Falling)      & 0.9\,V/\si{\micro\second}, 1.4\,V/\si{\micro\second} \\
Input Offset Voltage              & $-30$\,mV \\
Input Voltage Range               & $-1.5$\,V, 1.2\,V \\
Output Voltage Range              & $-1.5$\,V, 0.8\,V \\
Output Impedance                  & 230\,$\Omega$ \\
Dual Power Supply                 & $\pm$1.5\,V \\
Power Consumption                 & 71\,\si{\micro\watt} \\
Area                              & 0.0014\,mm$^2$ \\
\bottomrule
\end{tabular}
\end{table}

\begin{figure}[t!]
    \vspace{-3ex}
    \centering
    \includegraphics[width=1\columnwidth]{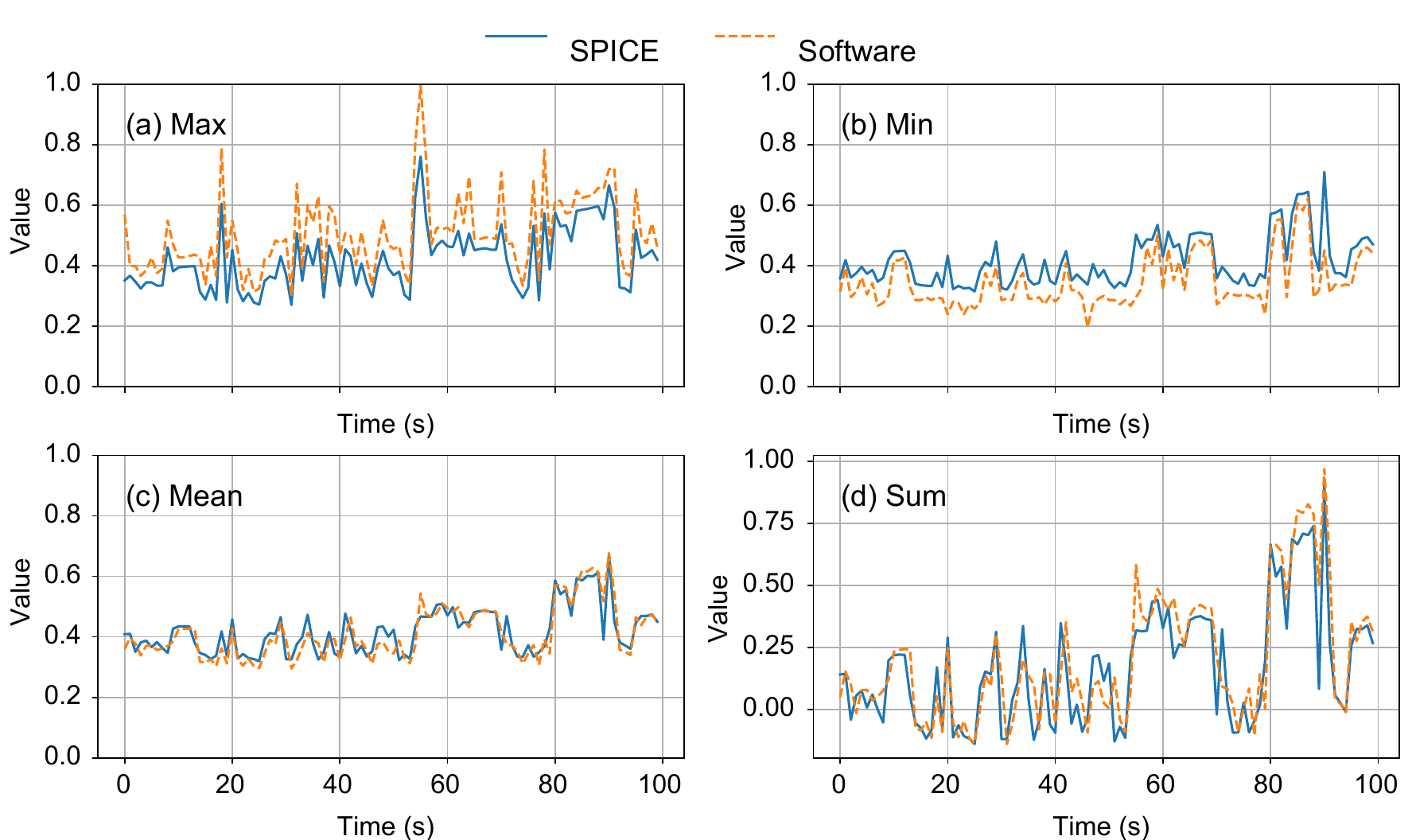}
\caption{Output of the Max/Min/Mean/Sum analog circuits (blue) versus ideal feature value from software (orange) for the SPD dataset and the accelerometer x-axis signal}

    \label{fig:Feature_validation}
    \vspace{-1ex}
\end{figure}



\begin{table}[t!]
\centering
\caption{Evaluation of analog feature extraction circuits}
\setlength{\tabcolsep}{6pt}
\begin{tabular}{l c c c}
\toprule
\multicolumn{1}{c}{\textbf{Feature}} & \multicolumn{1}{c}{\textbf{NMSE}} & \multicolumn{1}{c}{\textbf{Power (\si{\micro\watt})}} & \multicolumn{1}{c}{\textbf{Area (\si{\square\milli\meter})}} \\
\midrule
Max   & 0.005 & 0.44 & 0.0020 \\
Min   & 0.004 & 0.44 & 0.0020 \\
Mean  & 0.003 & 155  & 0.0084 \\
Sum   & 0.002 & 232  & 0.0099 \\
\bottomrule
\end{tabular}
\label{tab:feat_validation_hw}\vspace{-2ex}
\end{table}

The op-amp characterization results are summarized in Table~\ref{tab:opamp_results}, showing a good enough gain of $28.6$dB.
The power consumption of our SAR ADC is measured to be $81.4$$\mu$W and its area is as low as $0.02$mm$^2$.

Next, we evaluate our analog features as standalone statistical circuits. 
Fig.~\ref{fig:Feature_validation} presents a comparison between the SPICE analog output and the software reference value, for one sensor (accelerometer x-axis) from the SPD dataset.
\blue{For \emph{Sum}, to avoid early saturation, the circuit used a partial scaling by a representative factor \(N\); the residual factor was applied in software so the final trace reflects the intended sum for a fair SPICE--software comparison.
It is also adjusted to fall within a similar range to the other three features.}
As shown in Fig.~\ref{fig:Feature_validation}, the analog output closely follows the ideal software output.
Table~\ref{tab:feat_validation_hw} reports the accuracy and hardware measurements of our analog circuits as function approximators.
For the former, we consider the Normalized Mean Squared Error (NMSE) between our circuit outputs--obtained through SPICE simulations--and ideal software-computed features, averaged across the test set of all considered benchmarks (WESAD, SPD, and Stress-In-Nurses).
As shown, all features achieve low NMSE, indicating small absolute deviations.
In terms of hardware metrics, \emph{Max} and \emph{Min} share identical hardware overheads since they employ the same topology, differing only in diode orientation, whereas \emph{Mean} and \emph{Sum} incorporate an op-amp–based stage with its feedback/bias network, resulting in relatively higher power and area, as expected.
Indicatively, the reported area overheads are $97$\% lower compared to the respective digital feature extraction circuits.

\subsection{System Evaluation}
\label{sec:eval_system}
In this section, we evaluate our complete classification systems—comprising all components: analog feature extractors, ADC, and MLP classifier—obtained through our feature-to-classifier co-design.
We focus on three key metrics: accuracy, which defines the system’s performance; area, which is critical in FE applications due to the limited number of integrated devices~\cite{tahoori2025computing}; and energy per inference, which determines battery lifetime.
For reference, we compare our solutions against~\cite{afentaki2025islped}, where
a statistical-based feature selection method is used, combined with a brute-force exploration over the number of selected features, pruning sparsity, and weight quantization precision to identify area-efficient solutions.
Hereafter, we refer to SoA 4F as the solution from~\cite{afentaki2025islped} when considering only the \emph{min}, \emph{max}, \emph{mean}, and \emph{sum} features (as in our work) in their feature selection process, and to SoA AF as the respective solutions when all twelve statistical features (see Fig.\ref{fig:feature_histogram}) are considered in~\cite{afentaki2025islped}.
In~\cite{afentaki2025islped}, both feature extractors and MLPs are implemented in digital.
Table~\ref{tab:comparison_results} presents the comparative results for all approaches and datasets.

As shown in Table~\ref{tab:comparison_results}, our circuits deliver high classification accuracy with well-contained hardware costs.
Specifically, power consumption reaches only $20.3$ mW--well below the capabilities of existing printed batteries (e.g., a \SI{30}{\milli\watt} Molex printed battery)--while energy per inference remains below $1$ µJ.
The area ranges from $0.06$ mm$^2$ to $27.43$ mm$^2$, \textit{while achieving $73$\% to $89$\% classification accuracy and real-time monitoring within an accessible, mechanically flexible, and conformable healthcare wearable.}
Moreover, it is noteworthy that, despite any inaccuracies introduced by our analog feature extraction circuits, the classification accuracy remains within $3$\% of the purely software-based floating-point results.
The highest accuracy degradation is observed in SPD, most likely because it requires $21$ features, whereas the other datasets require fewer than $5$.
Finally, as shown in Table~\ref{tab:comparison_results}, unlike our motivation study, our co-design, combined with our area-efficient analog features, effectively reduces the feature extraction area to a negligible portion of the overall cost.

Compared to~\cite{afentaki2025islped}, our solutions achieve significantly higher accuracy (around $8$\% on average), despite the accuracy drop due to analog feature extraction, \blue{due to our integration of feature selection directly into \gls{mlp} training, whereas~\cite{afentaki2025islped}, as typically done~\cite{Shiyi:IoT2022:StressMonitoring}, employs a pre-training statistical feature selection to reduce the complexity of design space exploration.}
\blue{Note that the designs of~\cite{afentaki2025islped} are purely digital, achieving the expected software accuracy.}
Still, our solutions achieve higher accuracy than both SoA AF---demonstrating that limiting our solutions to the respective four selected analog features did not compromise accuracy---and SoA 4F, showing that incorporating our regularizer during training to minimize feature extraction cost likewise does not affect the achievable accuracy.
Moreover, our analog implementation of feature extraction allows for orders of magnitude lower area compared to digital features--of more than $600\times$--partly due to our effective feature selection scheme and due the area-efficiency of analog features.
Overall, our solutions achieve an average area reduction of $48\times$ compared to the most efficient solution of~\cite{afentaki2025islped}, while energy per inference is reduced by $70\times$, enabling significantly longer energy autonomy compared to purely digital implementations.

\begin{table}[t]
\setlength{\tabcolsep}{2pt}
\renewcommand{\arraystretch}{1.1}
\centering
\caption{Comparative in-depth analysis of our proposed approach against the state of the art~\cite{afentaki2025islped}}
\begin{threeparttable}
\begin{tabular}{
    l | 
    S[table-format=2.2]
    S[table-format=2.2]
    S[table-format=3.3]
    S[table-format=1.3]
    S[table-format=3.3]
    S[table-format=2.3]
    c
}
    \toprule
        Technique &
        {\thead{Software\\Accuracy\\(\%)}} & 
        {\thead{Circuit\\Accuracy\\(\%)}} &
        {\thead{Total\\Area\\(\si{\milli\meter^2})}} & 
        {\thead{Feature\\Area\\(\si{\milli\meter^2})}} &
        {\thead{Total\\Power\\(\si{\milli\watt})}} & 
        {\thead{Energy/\\Inference\\(\si{\micro\joule})}}
    \\
    \midrule
        \multicolumn{7}{c}{\textbf{WESAD}} \\ \midrule
        SoA AF\tnote{1}~\cite{afentaki2025islped} & 
        68.35 & 68.35 & 216.5 & 7.243 & 166.0 & 9.31
    \\
        SoA 4F\tnote{2}~\cite{afentaki2025islped} &
        73.95 & 73.95 & 241.3 & 3.581 & 178.7 & 10.5
    \\
        \textbf{Ours} &
        83.66 & 81.52 & 4.092 & 0.004 & 2.963 & 0.056
    \\
    \midrule
        \multicolumn{7}{c}{\textbf{Stress-In-Nurses}} \\ \midrule
        SoA AF\tnote{1}~\cite{afentaki2025islped} & 
        72.69 & 72.69 & 17.3 & 7.99 & 7.161 & 0.313
    \\
        SoA 4F\tnote{2}~\cite{afentaki2025islped} &
        74.46 & 74.46 & 66.43 & 3.579 & 30.62 & 2.5
    \\
        \textbf{Ours} &
        89.12 & 89.08 & 0.198 & 0.004 & 0.136 & 0.008

    \\
    \midrule
        \multicolumn{7}{c}{\textbf{SPD}} \\ \midrule
        SoA AF\tnote{1}~\cite{afentaki2025islped} & 
        67.60 & 67.60 & 105.5 & 7.971 & 73.03 & 3.83
    \\
        SoA 4F\tnote{2}~\cite{afentaki2025islped} &
        67.32 & 67.32 & 86.79 & 1.937 & 62.03 & 3.28
    \\
        \textbf{Ours} &
        73.42 & 70.41 & 27.57 & 0.078 & 20.3 & 0.763
    \\
    \bottomrule
\end{tabular}
\begin{tablenotes}[flushleft]
\blue{
\item[]\textsuperscript{1}Using all $12$ features from~\cite{afentaki2025islped}.
\textsuperscript{2}Using only \emph{min}, \emph{max}, \emph{mean}, and \emph{sum}.}
\end{tablenotes}
\end{threeparttable}\vspace{-3ex}
\label{tab:comparison_results}
\end{table}

\section{Conclusion}
\label{sec:conclusion}
\glspl{flexic} offer a compelling alternative to rigid silicon for wearable healthcare devices, enabling lightweight, conformable, and low-cost systems.
However, their large feature sizes and limited integration density impose strict area and power constraints that challenge full \gls{ml} implementations.
In this work, we introduced the first feature-to-classifier co-design framework for mixed-signal flexible systems, combining custom analog feature extractors with an optimized SAR ADC and a hardware-aware in-training feature selection.
Across multiple healthcare benchmarks, our approach achieves real-time operation with high accuracy, energy consumption below $1$uJ per inference, and practical area requirements.

\section*{Acknowledgment}
This work is partially supported by the \blue{European Research Council (ERC) (Grant No. 101052764)} and co-funded by the H.F.R.I call “Basic Research Financing (Horizontal support of all Sciences)” under the National Recovery and Resilience Plan “Greece 2.0” (H.F.R.I. Project Number: 17048).

\clearpage
\balance
{\linespread{0.98}\selectfont

}

\end{document}